\documentclass{emulateapj}

\usepackage{float}
\usepackage{amsmath}
\usepackage{epsfig,floatflt}
\usepackage{subfigure}

\begin{document}

\title{The joint large-scale foreground---CMB posteriors of the 3-year WMAP data}

\author{H.\ K.\ Eriksen\altaffilmark{1,2,3}, C.
  Dickinson\altaffilmark{4}, J. B. Jewell\altaffilmark{4}, A. J.
  Banday\altaffilmark{5}, K. M. G\'{o}rski\altaffilmark{4,7}, C. R.
  Lawrence\altaffilmark{4}}

\altaffiltext{1}{email: h.k.k.eriksen@astro.uio.no}

\altaffiltext{2}{Institute of Theoretical Astrophysics, University of
Oslo, P.O.\ Box 1029 Blindern, N-0315 Oslo, Norway}

\altaffiltext{3}{Centre of
Mathematics for Applications, University of Oslo, P.O.\ Box 1053
Blindern, N-0316 Oslo}

\altaffiltext{4}{Jet Propulsion Laboratory, California Institute
of Technology, Pasadena CA 91109} 

\altaffiltext{5}{Max-Planck-Institut f\"ur Astrophysik,
Karl-Schwarzschild-Str.\ 1, Postfach 1317, D-85741 Garching bei
M\"unchen, Germany}

\altaffiltext{6}{Warsaw University Observatory, Aleje Ujazdowskie 4, 00-478 Warszawa,
  Poland}

\date{Received - / Accepted -}

\begin{abstract}
  Using a Gibbs sampling algorithm for joint CMB estimation and
  component separation, we compute the large-scale CMB and foreground
  posteriors of the 3-yr WMAP temperature data.  Our parametric data
  model includes the cosmological CMB signal and instrumental noise, a
  single power law foreground component with free amplitude and
  spectral index for each pixel, a thermal dust template with a single
  free overall amplitude, and free monopoles and dipoles at each
  frequency. This simple model yields a surprisingly good fit to the
  data over the full frequency range from 23 to 94\,GHz.  We obtain a
  new estimate of the CMB sky signal and power spectrum, and a new
  foreground model, including a measurement of the effective spectral
  index over the high-latitude sky.  A particularly significant result
  is the detection of a common spurious offset in all frequency bands
  of $\sim-13\,\mu\textrm{K}$, as well as a dipole in the V-band data.
  Correcting for these is essential when determining the effective
  spectral index of the foregrounds. We find that our new foreground
  model is in good agreement with template-based model presented by
  the WMAP team, but not with their MEM reconstruction. We believe the
  latter may be at least partially compromised by the residual offsets
  and dipoles in the data.  Fortunately, the CMB power spectrum is not
  significantly affected by these issues, as our new spectrum is in
  excellent agreement with that published by the WMAP team. The
  corresponding cosmological parameters are also virtually unchanged.
\end{abstract}

\keywords{cosmic microwave background --- cosmology: observations --- 
methods: numerical}

\maketitle

\section{Introduction}

A major challenge in CMB research is component separation, which can
be summarized in two questions. First, how can we separate reliably
the valuable cosmological signal from confusing foreground emission?
Second, how can we propagate accurately the errors induced by this
process through to the final analysis products, such as the CMB power
spectrum and cosmological parameters?

During the last few years, a new analysis framework capable of
addressing these issues in a statistically consistent approach has
been developed. This framework is Bayesian in nature, and depends
critically on the Gibbs sampling algorithm as its main computational
engine. The pioneering ideas were described by
\citet{jewell:2004} and \citet{wandelt:2004}, and later implemented
for modern CMB data sets for temperature and polarization by
\citet{eriksen:2004a} and \citet{larson:2007}, respectively.
Applications to the 1-yr and 3-yr WMAP data \citep{bennett:2003a,
  hinshaw:2007, page:2007} were described by \citet{odwyer:2004}, and
\citet{eriksen:2006,eriksen:2007a}. These papers mainly focused on the
cosmological CMB signal, and adopted the foreground corrected data
provided by the WMAP team.

Recently this algorithm was extended to include internal component
separation capabilities by \citet{eriksen:2007b}.  Using very general
parameterizations of the foreground components, this method produces
the full joint and exact foreground-CMB posterior, and therefore
allows us both to estimate each component separately through
marginalized statistics and to propagate the foreground uncertainties
through to the final CMB products. The implementation of this
algorithm used in this paper is called ``Commander'', and is a direct
descendant of the code presented by \citet{eriksen:2004a}.

In this Letter, we apply the method to the 3-yr WMAP temperature
observations \citep{hinshaw:2007}.  This data set, with five frequency
bands, allows only very limited foreground models; however, the
analysis provides a powerful demonstration of the capabilities of the
method. For a comprehensive analysis of a controlled simulation with
identical properties to this data set, see \citet{eriksen:2007b}.

\section{Data}
\label{sec:data}

We consider the 3-yr WMAP temperature data, provided on
Lambda\footnote{http://lambda.gsfc.nasa.gov} in the form of sky maps
from ten ``differencing assemblies'' covering the frequency range
between 23 and 94\,GHz. Since our current implementation of the Gibbs
foreground sampler can only handle sky maps with identical beam
response \citep{eriksen:2007b}, we downgrade each of these maps to a
common resolution of $3^\circ$ FWHM and repixelize at a
HEALPix\footnote{http://healpix.jpl.nasa.gov} resolution of
$N_{\textrm{side}}=64$, corresponding to a pixel size of $55'$. These
ten maps are then co-added by frequency into five single frequency
band maps at 23, 33, 41, 64 and 94\,GHz (K, Ka, Q, V and W-bands,
respectively).

The power from the instrumental noise is less than 1\% of the CMB
signal at $\ell=50$ in the V- and W-bands, and less than 2\% at
$\ell=100$ \citep{eriksen:2007b}. To regularize the noise covariance
matrix at high spatial frequencies, we added $2\,\mu\textrm{K}$ per
$3^\circ$ pixel of uniform white noise.  This noise is insignificant
at low multipoles, but dominates the signal near the spherical
harmonic truncation limit of $\ell_{\textrm{max}}=150$.  We then have
five frequency maps at a common resolution of $3^{\circ}$ FWHM, with
signal-to-noise ratio of unity at $\ell \sim 120$, and strongly
bandwidth limited at $\ell_{\textrm{max}} = 150$.

We choose to include such high $l$'s in the analysis for two reasons.
First, our main goal is an accurate approximation of the CMB
likelihood at $\ell \le 50$. In order to ensure that the degradation
process (i.e., smoothing and noise addition) does not significantly
affect these multipoles, it is necessary to go well beyond $\ell \sim
80$--100. Second, significant information on the spatial distribution
of foregrounds is obtained by going to higher resolution.

The cost of this treatment of the noise is high $\chi^2$ values in the
ecliptic plane \citep{eriksen:2007b}, where the instrumental noise is
higher because of WMAP's scanning strategy. However, since these high
$\chi^2$ values are caused by unmodelled smoothed, random, white
noise, they do not indicate a short-coming of the signal model, but
only a slight under-estimation of the statistical errors on small
angular scales.  This has been confirmed by otherwise identical
analyses at both $4^{\circ}$ and $6^{\circ}$ FWHM.  We present the
$3^{\circ}$ FHWM case here as a compromise between angular resolution
and accuracy of the noise model.  This issue will further suppressed
with additional years of WMAP observations and, eventually,
high-sensitivity Planck maps.

We impose the base WMAP Kp2 sky cut \citep{bennett:2003b} on the data,
but not the point source cuts.  The base mask is downgraded from its
native $N_{\textrm{side}}=512$ resolution to $N_{\textrm{side}} = 64$
by excluding all low-resolution pixels for which any one of its
sub-pixels is excluded by the high-resolution mask. A total of 42\,081
pixels are included, or 85.6\% of the sky.

The frequency bandpass of each map is modelled as a top-hat
function, and implemented in terms of effective frequency as a
function of spectral index as described by \citet{eriksen:2006}. The
frequency specifications of the WMAP radiometers are given by
\citet{jarosik:2003}.

\section{Model and methods}
\label{sec:model}

We adopt the following simple parametric model $T_{\nu}(p)$ for the
observed signal (measured in thermodynamic temperatures) at frequency
$\nu$ and pixel $p$,
\begin{equation}
\begin{split}
  T_{\nu}(p) &= s(p) + m^{0}_{\nu} + \sum_{i=1}^{3} m^i_{\nu}
  \left[\hat{\mathbf{e}}_i\cdot\hat{\mathbf{n}}(p)\right] + \\+
  b&\left[t(p) a(\nu) \left(\frac{\nu}{\nu_0^{\textrm{dust}}}\right)^{1.7}\right] + f(p) a(\nu)
  \left(\frac{\nu}{\nu_0}\right)^{\beta(p)}.
\end{split}
\end{equation}
The first term is the cosmic CMB signal, characterized by a
frequency-independent spectrum and a covariance matrix in spherical
harmonic space given by the power spectrum, $\left<a^*_{\ell m}
  a_{\ell' m'}\right> = C_{\ell} \delta_{\ell \ell'} \delta_{m m'}$,
where $s(p) = \sum_{\ell, m} a_{\ell} Y_{\ell m}(p)$. The second and
third terms denote a free monopole and three dipole amplitudes at each
frequency. We use the standard Cartesian basis vectors projected on
the sky, $\{\mathbf{1}, \mathbf{x}, \mathbf{y}, \mathbf{z}\}$, as
basis functions for these four modes. The fourth term represents a
template-based dust model scaled by a spectral index of $\beta=1.7$,
in which $t(p)$ is the dust template (FDS) of \citet{finkbeiner:1999},
evaluated at $\nu_{0}^{\textrm{dust}}=$ 94\,GHz, and $a(t)$ is the
conversion factor between antenna and thermodynamic temperatures. The
last term is a single foreground component with a free amplitude
$f(p)$ and spectral index $\beta(p)$ at each pixel. The reference
frequency for this component is $\nu_0 = 23$\,GHz.

The free parameters in this model are: 1)~spherical harmonic
coefficients $a_{\ell m}$ of the CMB amplitude $s$; 2)~CMB power
spectrum coefficients $C_{\ell}$; 3)~monopole and dipole amplitudes at
each band; 4)~the amplitude of the dust template; and 5)~amplitudes
and spectral indices of the pixel foreground component.

The WMAP data do not have sufficient power to constrain this simple
completely by themselves, as there is a very strong degeneracy between
the foreground component amplitudes at each pixel and the free
monopole and dipole coefficients at each band \citep{eriksen:2007b}.
For this reason, we introduce two priors in addition to the Jeffereys'
ignorance prior discussed by \citet{eriksen:2007b}. First, we impose a
Gaussian prior on the spectral indices of $\beta \sim -3\pm0.3$: A
direct fit of the 408 MHz template \citep{haslam:1981} to the WMAP
K-band data for Kp2 sky coverage implies an index of -3
\citep{davies:2006}, and \citet{davies:1996} determined a typical
range for high latitude spectral indices between 408~MHz and 1420~MHz
of -2.8 to -3.2. Note that this prior has a noticeable effect only at
high Galactic latitudes, where the absolute foreground amplitude is
low. At low galactic latitudes, the data dominate the prior by up to a
factor of $\sim50$, and any potential bias in the near-plane free-free
regions is negligible.

Second, we impose an implicit spectral index orthogonality prior on
the monopole and dipole coefficients, as described by
\citet{eriksen:2007b}, projecting out the frequency component of these
coefficients that matches the free spectral index map, thus
effectively determining the zero-level of the foreground amplitude
map.  We also tried an alternative approach, first estimating the Q,
V, and W-band monopole and dipole coefficients separately given a
crude estimate of the spectral index map, and then estimating all
other parameters given these coefficients.  Results were very similar.
Thus, the two priors adopted in this analysis have a very weak effect
on all main results.

Having defined our model and priors, we map out the joint posterior
distribution using the foreground Gibbs sampler described by
\citet{eriksen:2007b}.  We refer the interested reader to that paper
for full details of the algorithm, and for a comprehensive analysis of
a realistic simulation corresponding to the same data and model used
in this Letter.

Finally, we estimate a new set of cosmological parameters within the
standard $\Lambda$CDM model. For this analysis, we follow the approach
of \citet{eriksen:2007a}, and replace the low-$\ell$ part of the WMAP
likelihood with a new Blackwell-Rao Gibbs-based estimator
\citep{chu:2005}. No ancillary data sets beyond the 3-yr WMAP
temperature and polarization data are included in the analysis. The
CosmoMC code \citep{lewis:2002} is used as the main MCMC engine.

\section{Results}
\label{sec:results}

\begin{figure}[t]
\mbox{\epsfig{figure=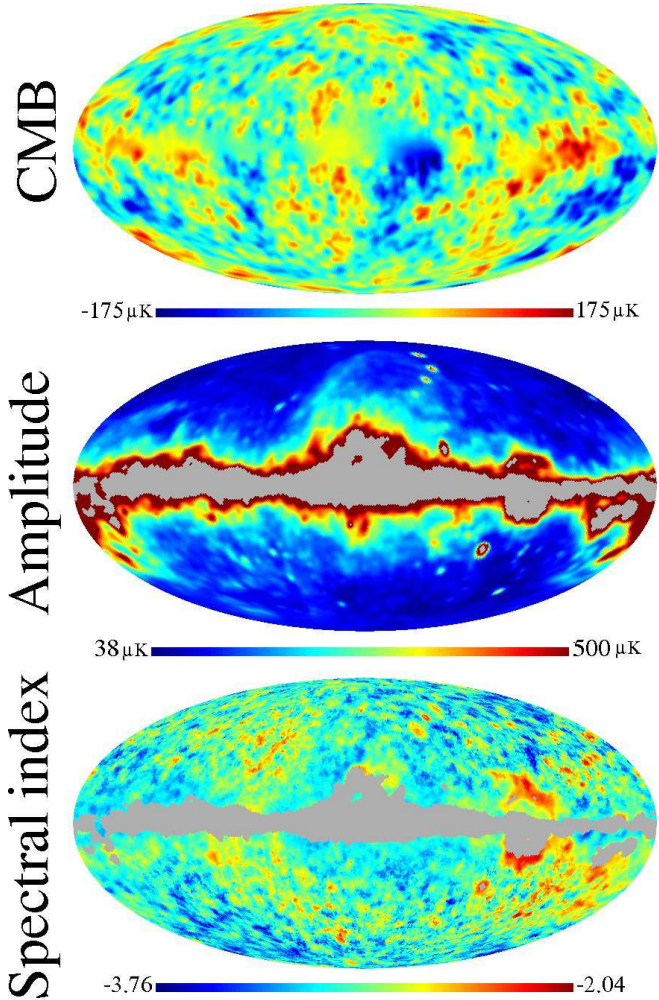,width=\linewidth,clip=}}
\caption{Marginal posterior mean maps in Galactic coordinates. Rows
  from top to bottom show the CMB reconstruction, the foreground
  amplitude, and the foreground spectral index, respectively.}
\label{fig:sky_maps}
\end{figure}

Figure \ref{fig:sky_maps} shows the marginal posterior mean maps for
the CMB sky signal, the foreground amplitude and the foreground
spectral index.  Table \ref{tab:monopole} gives the corresponding
results for the monopole and dipole coefficients for each frequency
band.  The FDS dust template amplitude relative to 94\,GHz and an
assumed spectral index of $\beta=1.7$ is $b=0.917\pm0.003$. The CMB
power spectrum is discussed separately below.

\begin{deluxetable}{lcccc}
\tablewidth{0pt} 
\tabletypesize{\small} 
\tablecaption{Monopole and dipole posterior statistics\label{tab:monopole}}
\tablecolumns{5}
\tablehead{ & Monopole & Dipole X & Dipole Y & Dipole Z \\
Band & ($\mu\textrm{K}$) & ($\mu\textrm{K}$)& ($\mu\textrm{K}$)&
 ($\mu\textrm{K}$)
}

\startdata

K-band    & $-11.8 \pm 0.5$ & $1.7 \pm 1.2$ & $-2.2 \pm 0.8$ & $\phm{-}2.2  \pm 0.1$ \\
Ka-band   & $-16.6 \pm 0.5$ & $0.9 \pm 1.2$ & $\phm{-}0.9 \pm 0.8$ & $-1.3 \pm 0.1$ \\
Q-band    & $-12.8 \pm 0.5$ & $1.9\pm 1.2$  & $-0.9 \pm 0.8$ & $\phm{-}0.4  \pm 0.1$\\ 
V-band    & $-11.1 \pm 0.5$ & $ 1.6\pm 1.2$ & $-3.9 \pm 0.8$ & $\phm{-}4.0  \pm 0.1$ \\ 
W-band    & $-12.6 \pm 0.5$ & $ 1.7\pm 1.2$ & $-0.9 \pm 0.8$ & $\phm{-}\,\,\,1.0  \pm 0.1$

\enddata
\tablecomments{Means and standard deviations of the marginal monopole and dipole posteriors.}
\end{deluxetable}

\begin{figure}[t]
\mbox{\epsfig{figure=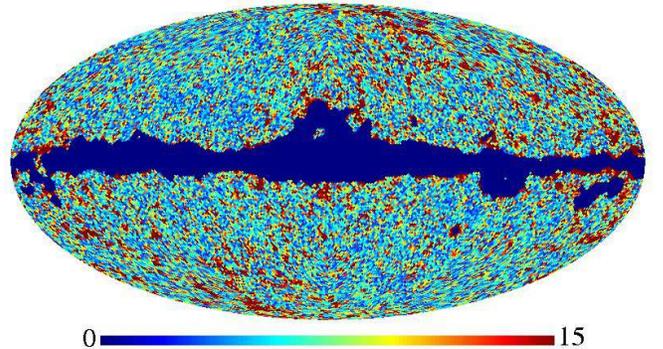,width=\linewidth,clip=}}
\caption{Mean $\chi^2$ map computed over posterior samples in Galactic
  coordinates. A value of $\chi^2 = 15$  is high at the 99\% significance
  level.}
\label{fig:chisq}
\end{figure}

Figure \ref{fig:chisq} shows the average $\chi^2$ computed for each
Gibbs sample.  A $\chi^2$ value exceeding $\chi^2 = 15$ corresponds to
rejection of the model in that pixel at 99\% statistical significance.
Two features are clearly visible in this plot. First, the ecliptic
plane, or rather, WMAP's scanning strategy, is clearly visible, and
this is mainly due to the unmodelled smoothed noise component at high
$\ell$'s, as discussed in Section \ref{sec:data}.

Second, there are clearly visible structures near the Galactic plane,
and in particular around regions with known high free-free emission
(e.g., the Orion and Ophiucus regions). This is likely due to the fact
that a single power-law is not a good approximation to the sum of many
foregrounds with comparable amplitudes.  Thus, we have clear evidence
of modelling errors in this solution, and we therefore strongly
emphasize that the quoted error bars presented in this paper include
formal statistical errors only, and not systematic, model-dependent
uncertainties. These issues are discussed in depth by
\citet{eriksen:2007b}, who find similar behaviour for a controlled,
simulated data set.

Although $\chi^2$ is high, the CMB solution obtained is evidently
good.  First of all, in the ecliptic poles, where the WMAP
instrumental noise is suppressed by the scanning strategy, the
$\chi^2$ distribution is essentially perfect. This implies that the
signal model as such is adequate at high latitudes.  Second, the CMB
map is virtually without signatures of residual foregrounds. (Note
that the signal inside the Galactic plane is partially reconstructed
using high-latitude information and the assumption of isotropy. The
signal on scales smaller than the mask size is washed out because it
is not possible to predict these from higher latitudes.)  Finally, the
foreground amplitude and spectral index maps correlate very well with
known templates of synchrotron and free-free emission. For instance,
the spectral index near the Gum nebula and near the Vela regions are
close to $\beta=-2.1$, as expected for free-free emission. 

The single most surprising aspects of the solution are the monopole
and, possibly, the V-band dipole coefficients, listed in Table
\ref{tab:monopole}. Most notably, there is a strong detection of a
roughly $\-13 \mu\textrm{K}$ offset common to all frequency bands.
Formally speaking, these offset values are only optimal within the
current model; however, this type of signal is not degenerate with any
other component in the model.  Further, we have attempted to fit
several other models assuming no offsets at one or more bands.  These
all result in strong, visible residuals in the CMB map, and
considerably higher $\chi^2$ values overall.  Finally, very similar
results have been obtained by other researchers\footnote{See
  discussion lead by P.\ Leahy at
  http://cosmocoffee.info/viewtopic.php?t=631.} through other methods,
although, to our knowledge, these results have not yet been published
in the literature. We therefore believe that the monopole and dipole
coefficients presented here are more optimal even in an absolute sense
than those obtained by WMAP based on a cosecant fit to a plane
parallel galaxy model \citep{bennett:2003b}.

\begin{figure}[t]
\mbox{\epsfig{figure=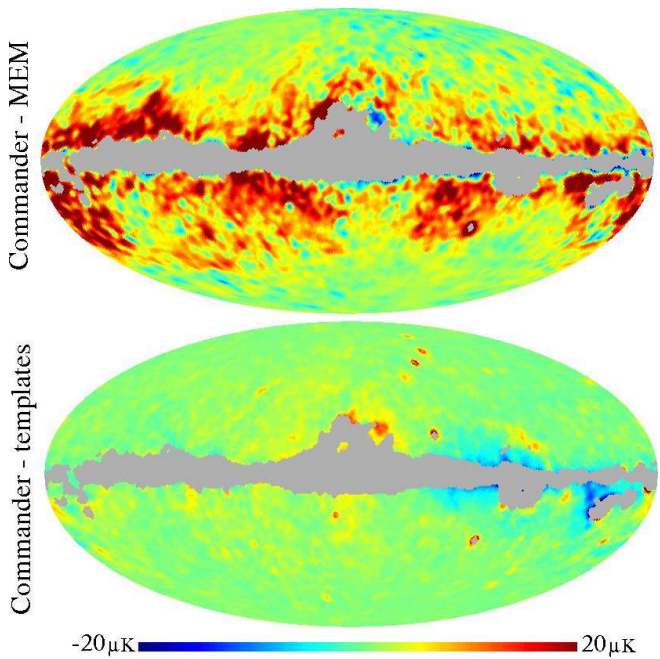,width=\linewidth,clip=}}
\caption{Difference maps of the total ``Commander'' W-band foreground
  model with the WMAP MEM model (top panel) and with the WMAP template
  fit model (bottom panel).}
\label{fig:diff_maps}
\end{figure}

Figure \ref{fig:diff_maps} shows the difference maps between the
Commander W-band foreground model and the MEM and template-based
foreground models of \citet{hinshaw:2007}.  Clearly, our foreground
model agrees surprisingly well with the simple template fits, but not
with the MEM solution. One possible explanation for this is that
although the MEM approach of \citet{hinshaw:2007} does attempt to
estimate spectral indices for each pixel, it does not include monopole
or dipole components in its model. Therefore, the MEM solution is
plausibly compromised by the non-zero offset detected here, at least
in part. In addition, the MEM method could be biased by the initial
subtraction of the (foreground contaminated) ILC estimate of the CMB
anisotropy from the frequency maps, and the use of the 408 MHz data as
a prior. Conversely, Commander could be compromised to some extent by
the use of power law spectral indices for the combined low frequency
foreground component. Nevertheless, the difference is surprising given
that the W-band foreground is expected to be mostly comprised of
thermal dust emission.

\begin{figure}[t]
\mbox{\epsfig{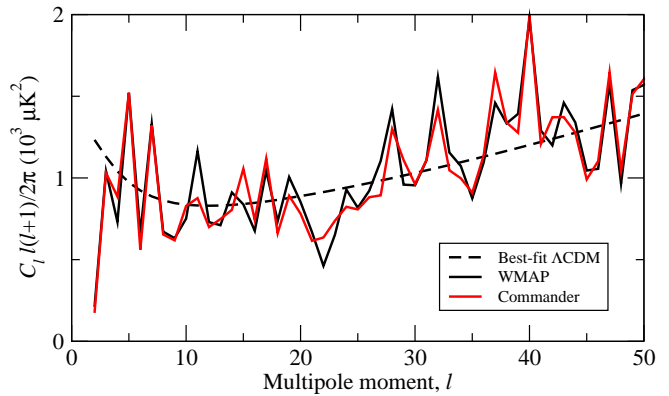}}
\caption{The CMB temperature power spectra obtained
  in this paper (red) and that by the WMAP team (black). The best-fit
  $\Lambda$CDM model spectrum of \citet{spergel:2007} is shown as a
  dashed line.}
\label{fig:powspec}
\end{figure}

The marginal maximum posterior CMB power spectrum is shown in Figure
\ref{fig:powspec}, together with the maximum-likelihood/
pseudo-$C_{\ell}$ hybrid spectrum computed by the WMAP team.  Perhaps
the most notable difference is in the $\ell=21$ multipole, which looks
anomalous in the WMAP spectrum, as noted by other authors.

The cosmological parameters corresponding to the Commander spectrum
for a standard six-parameter $\Lambda$CDM model are
$\Omega_{\textrm{b}}\,h^2 =0.0222 \pm 0.0007$, $\Omega_{\textrm{m}} =
0.243 \pm 0.036$, $\log(10^{10}A_{\textrm{s}}) = 3.027 \pm 0.068$, $h
= 0.730 \pm 0.032$ and $\tau = 0.089 \pm 0.030$. Corresponding values
for the unmodified WMAP likelihood are $\Omega_{\textrm{b}}\,h^2
=0.0221 \pm 0.0007$, $\Omega_{\textrm{m}} = 0.242 \pm 0.035$,
$\log(10^{10}A_{\textrm{s}}) = 3.030 \pm 0.068$, $h = 0.730 \pm 0.032$
and $\tau = 0.091 \pm 0.030$. These values refer to marginal means and
standard deviations.

Clearly, the agreement between the two sets of results is excellent,
and this provides a strong confirmation of the WMAP results: At the
level of precision of the WMAP experiment, details in the foreground
model used for foreground correction appear to have only a minor
impact on the CMB temperature power spectrum. 

\section{Conclusions}
\label{sec:conclusions}

We have presented the first exact Bayesian joint foreground-CMB
analysis of the 3-yr WMAP data. We have established a new estimate of
both the CMB sky signal and the power spectrum, a detailed foreground
model consisting of a foreground amplitude and spectral index map and
a dust template amplitude, and also provided new estimates of the
residual monopole and dipole coefficients in the WMAP data.

The detection of significant non-zero offsets in the WMAP data is the
new result of the greatest immediate importance for the CMB community.
These new monopole and dipole estimates could have a significant
impact on several previously published results, especially those
concerning the foreground composition in the WMAP data. For example,
our foreground model is in excellent agreement with the simple
template fits presented by \citet{hinshaw:2007}, but not with their
MEM reconstruction.

Taking a longer perspective, the most important aspect of this
analysis is a demonstration of feasibility of exact and joint
foreground-CMB analysis.  This will be essential for Planck, whose
high sensitivity and angular resolution demand more accurate
foreground separation than \hbox{WMAP}. Considering the flexibility,
power, and accuracy of the method employed in this paper, together
with its unique capabilities for propagating uncertainties accurately
all the way from the postulated foreground model to cosmological
parameters, we believe that this should be the baseline analysis
strategy for Planck on large angular scales, say $\ell \lesssim 200$.

All results presented in this paper and the basic Gibbs samples are
available at http://www.astro.uio.no/$\sim$hke.

\begin{acknowledgements}
  We acknowledge use of the HEALPix software \citep{gorski:2005} and
  analysis package for deriving the results in this paper. We
  acknowledge use of the Legacy Archive for Microwave Background Data
  Analysis (LAMBDA). This work was partially performed at the Jet
  Propulsion Laboratory, California Institute of Technology, under a
  contract with the National Aeronautics and Space Administration. HKE
  acknowledges financial support from the Research Council of Norway.
\end{acknowledgements}

\end{document}